\documentclass[preprint2]{emulateapj}
\usepackage{natbib}
\usepackage{hyperref}
\usepackage{breakurl}
\bibpunct[; ]{(}{)}{,}{a}{}{,}

\newcommand{\flux}{erg~cm$^{-2}$~s$^{-1}$}

\usepackage{color}

\shorttitle{Lyman Continuum Observations of Solar Flares Using SDO/EVE}

\shortauthors{Machado, Milligan \& Sim\~{o}es}

\begin{document}

\title{Lyman Continuum Observations of Solar Flares Using SDO/EVE}

\notetoeditor{the contact email is ryan.milligan@glasgow.ac.uk and is the only one which should appear on the journal version}

\author{Marcos E. Machado\altaffilmark{1}, Ryan O. Milligan\altaffilmark{2,3,4}, \& Paulo J. A. Sim\~{o}es\altaffilmark{2}}

\altaffiltext{1}{Comisi\'{o}n Nacional de Actividades Espaciales (CONAE) Paseo Col\'{o}n 751, Buenos Aires C1063ACH Buenos Aires, Argentina}
\altaffiltext{2}{SUPA School of Physics and Astronomy, University of Glasgow, Glasgow, G12 8QQ, UK}
\altaffiltext{3}{Solar Physics Laboratory (Code 671), Heliophysics Science Division, NASA Goddard Space Flight Center, Greenbelt, MD 20771, USA}
\altaffiltext{4}{Department of Physics Catholic University of America, 620 Michigan Avenue, Northeast, Washington, DC 20064, USA}

\begin{abstract}
The {\em Extreme ultraviolet Variability Experiment} was designed to observe the Sun-as-a-star in the extreme ultraviolet; a wavelength range that has remained spectrally unresolved for many years. It has provided a wealth of data on solar flares, perhaps most uniquely, on the Lyman spectrum of hydrogen at high cadence and moderate spectral resolution. In this paper we concentrate on the analysis of Lyman continuum (LyC) observations and its temporal evolution in a sample of six major solar flares. By fitting both the pre-flare and flare excess spectra with a blackbody function we show that the color temperature derived from the slope of LyC reveals temperatures in excess of 10$^{4}$~K in the six events studied; an increase of a few thousand Kelvin above quiet-Sun values (typically $\sim$8000--9500~K). This was found to be as high as 17000~K for the 2017 September 6 X9.3 flare. Using these temperature values, and assuming a flaring area of 10$^{18}$~cm$^{2}$, estimates of the departure coefficient of hydrogen, $b_1$, were calculated. It was found that $b_1$ decreased from 10$^{2}$--10$^{3}$ in the quiet-Sun, to around unity during the flares. This implies that LyC is optically thick and formed in local thermodynamic equilibrium during flares. It also emanates from a relatively thin ($\lesssim$100~km) shell formed at deeper, denser layers than in the quiescent solar atmosphere. We show that in terms of temporal coverage and resolution, EVE gives a more comprehensive picture of the response of the chromosphere to the flare energy input with respect to those of the {\em Skylab/Harvard College Observatory} spatially resolved observations of the 1970's.
\end{abstract}

\keywords{Sun: activity --- Sun: chromosphere --- Sun: flares --- Sun: UV radiation}

\section{INTRODUCTION AND BACKGROUND}
\label{intro}
The {\em Extreme ultraviolet Variability Experiment} (EVE; \citealt{wood12}) was launched aboard the {\em Solar Dynamics Observatory} (SDO; \citealt{pesn12}) spacecraft on 2010 February 11, with the mission to study the Sun and its variability in the extreme ultraviolet (EUV) spectral range with high sensitivity and continuous coverage. It has since provided near-continuous spectral information at high cadence, including the Lyman continuum (LyC), Lyman lines, and other lines formed at chromospheric and transition region temperatures. Its full-Sun coverage ensures that flares are always observed when the instrument is exposed to the visible solar disk. EVE thus provides spatially-integrated EUV spectra of the most intense flaring regions, allowing the study of the evolution of physical parameters that contain valuable information on chromospheric energy deposition and transport. 

LyC is formed at the base of the transition region, the thin interface where the temperature rises from the cooler, denser layers of the chromospheres and photosphere ($T \le 10^4$~K) to the $>10^6$~K corona. More specifically, the bulk of the continuum emission comes from the very top of the chromosphere (see e.g. \citealt{vern73,vern76,vern81}; VAL, \citealt{avre08}), and is therefore subject to any change in the energy balance in these structures. In physical terms it is due to the recombination of free electrons to the first level of ambient hydrogen nuclei, and since at chromospheric densities the electron thermalization occurs very rapidly, and even more so in higher densities during flares, its spectrum reflects the ambient temperature. Its study is thus a very valuable tool as a probe to study the response of the chromosphere--transition region to the energy input due to a flare, and within timescales commensurate to those of its impulsive energy release.

The first comprehensive study of LyC observations of solar flares was performed 40 years ago by \cite{mach78}. These data, from the {\em Harvard College Observatory EUV spectroheliometer} on {\em Skylab}, combined with observations of other spectral features, were used to calculate semi-empirical chromospheric and photospheric flare models \citep{mach80,avre86,maua90}. The studies of \cite{mach78} showed significant increases (factors of 10 to 100) in LyC intensity compared to quiet-Sun values, while generally showing a continuum color temperature, $T_c \sim$ 8000--8500~K, similar to that observed in the quiet-Sun, with a marginally observed tendency to higher values in stronger events. Detailed modeling \citep{mach80,avre86} subsequently confirmed these results showing that the large intensity increases were associated with high densities at the layer where the observed head of the continuum (912~\AA) was formed, leading to an increased collisional coupling to the local temperature. As shown by \cite{noye70} the continuum radiance, $I_\lambda$, can be written in an Eddington-Barbier approximation as:
\begin{equation}
I_\lambda (\mu)  \approx  S_\lambda (\tau=\mu) = \frac{B_\lambda(T)}{b_1},
\label{eqn:source}
\end{equation}
where $B_\lambda(T)$ is the Planck function,
\begin{equation}
B_{\lambda}(T) = \frac{2hc^2}{\lambda^5} \frac{1}{\exp(\frac{hc}{\lambda k_B T})-1},
\label{eqn:planck}
\end{equation} 
and $b_1$ is the non local thermodynamic equilibrium (NLTE) departure coefficient. Since $b_1$ (the departure coefficient of the first hydrogen level, =$n_{1}/n_{1}^{*}$, where $n_{1}^{*}$ is the LTE population) is on the order of $10^{2}$--$10^{3}$ in the quiet-Sun, it follows that under the increased collisional coupling conditions due to higher electron densities in flares ($n_e \sim 10^{13}$~cm$^{-3}$), $b_1$ becomes closer to unity and an equivalent $I_\lambda$ increase ensues even for similar temperatures (cf. Equation~\ref{eqn:source}). Under these conditions, the measured color temperature, $T_c$, approximately equals the electron temperature, $T_e$. Similarly, LyC intensity can be increased by moving its formation layer to a higher temperature region, thereby increasing $B_\lambda(T)$.

Finally, \cite{mach78} also showed that $T_c$ determined at the shortest wavelengths ($\lambda \le$ 790~\AA) was found to be higher than those obtained closer to the continuum head. This indicated the presence of higher temperature plasmas in the Ly$\alpha$ forming region, with $T$ of the order of 10$^4$~K, which in the VAL models was represented by a 20000~K plateau that was later shown to be unnecessary. It is also worth noting that the \cite{mach78} flare sample referred in most cases to rather small and not too bright flares in H$\alpha$, as reported in Solar Geophysical Data.

Definitive observations of enhanced free-bound emission during solar flares have been scarce in the interim years as many modern space-based instruments have not had the sensitivity, wavelength coverage, or duty cycle required to capture such enhancements. One notable exception includes the detection of increased LyC emission during an X-class flare via scattered light in the {\em Solar Ultraviolet Measurements of Emitted Radiation} (SUMER; \citealt{wilh95}) instrument onboard the {\em Solar and Heliospheric Observatory} (SOHO) reported by \citet[see also \citealt{pare05} who calculated $T_c$ for prominences using LyC data from SUMER]{lema04}. There also appears to be only one reference in the literature that provides intensity maps of a flare in LyC emission, carried out using the Japanese sounding rocket S520-5CN \citep{hira85}. \cite{chri03} also reported an analogous increase in LyC emission during a stellar flare using data from the {\it Extreme Ultraviolet Explorer}. However, SDO/EVE now routinely detects changes in LyC emission during solar flares, the first of which was reported by \cite{mill12}. This allows for a more detailed exploration of the changes in physical conditions in the solar chromosphere in response to an injection of energy.

In this paper we present a detailed analysis of physical parameters in the solar chromosphere during six major solar flares using LyC data from SDO/EVE. Section~\ref{sec:data_anal} describes the analysis techniques that were utilized. The findings are presented in Section~\ref{sec:results}, while a discussion is presented in Section~\ref{sec:disc}. The conclusions are summarized in Section~\ref{sec:conc}.

\section{DATA ANALYSIS}
\label{sec:data_anal}

EVE was primarily designed to monitor changes in the Sun's EUV output over a broad range of wavelengths (and temperatures) on flaring timescales (10~s). The instrument comprises two Multiple EUV Grazing Spectrograph (MEGS) components that cover two distinct wavelength ranges: MEGS-A covers the 65--370~\AA\ range while MEGS-B spans 370--1050~\AA, which contains the Lyman continuum. MEGS-A operated uninterrupted until it suffered a power anomaly on 2014 May 26. The MEGS-B detector began to degrade soon after launch and so its exposure to the Sun was limited to three hours per day, and five minutes per hour, to sustain operations for as long as possible. After the loss of MEGS-A, the flight software on SDO was updated to command MEGS-B to respond to a flare trigger. Specifically, if the X-ray flux detected by the ESP channel of EVE exceeded the $\sim$M1 level, then MEGS-B would begin collecting data for the subsequent three hours. This now makes EVE a dedicated flare-hunting mission, rather than exposing for an arbitrary time period per day in the hopes that a flare will occur.

\begin{figure}[!t]
\begin{center}
\includegraphics[width=0.5\textwidth]{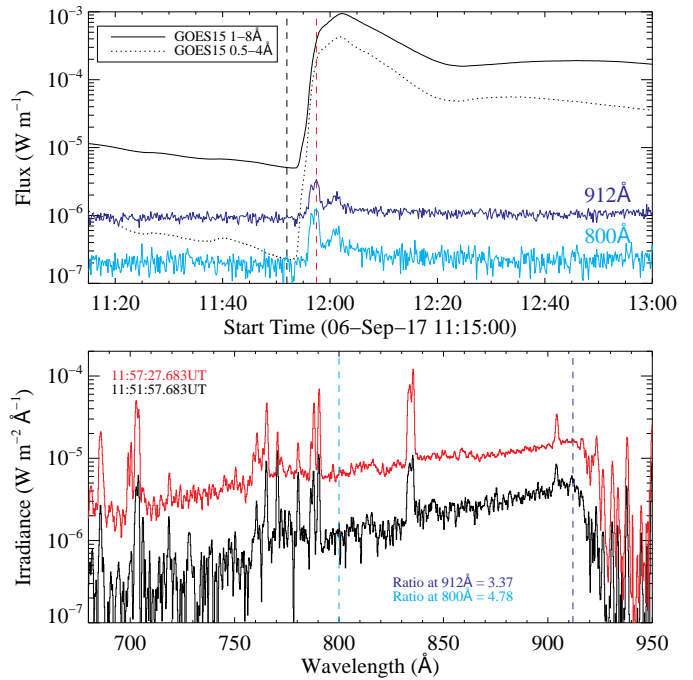}
\caption{Top panel: GOES soft X-ray lightcurves (solid and dotted curves) for the 2017 September 6 X9.3 flare, along with the time profiles of LyC emission at 800~\AA\ (cyan) and 912~\AA\ (blue). The vertical dashed black and red lines denote the times of the pre-flare and impulsive phase spectra in the bottom panels, respectively. Bottom panel: SDO/EVE MEGS-B pre-flare (black) and impulsive phase (red) spectra from 680--950~\AA. The vertical dashed cyan and blue lines denote the wavelength values of the time profiles in the top panel.}
\label{fig:lyc_head_ratio}
\end{center}
\end{figure}

\begin{figure*}[!t]
\begin{center}
\includegraphics[width=\textwidth]{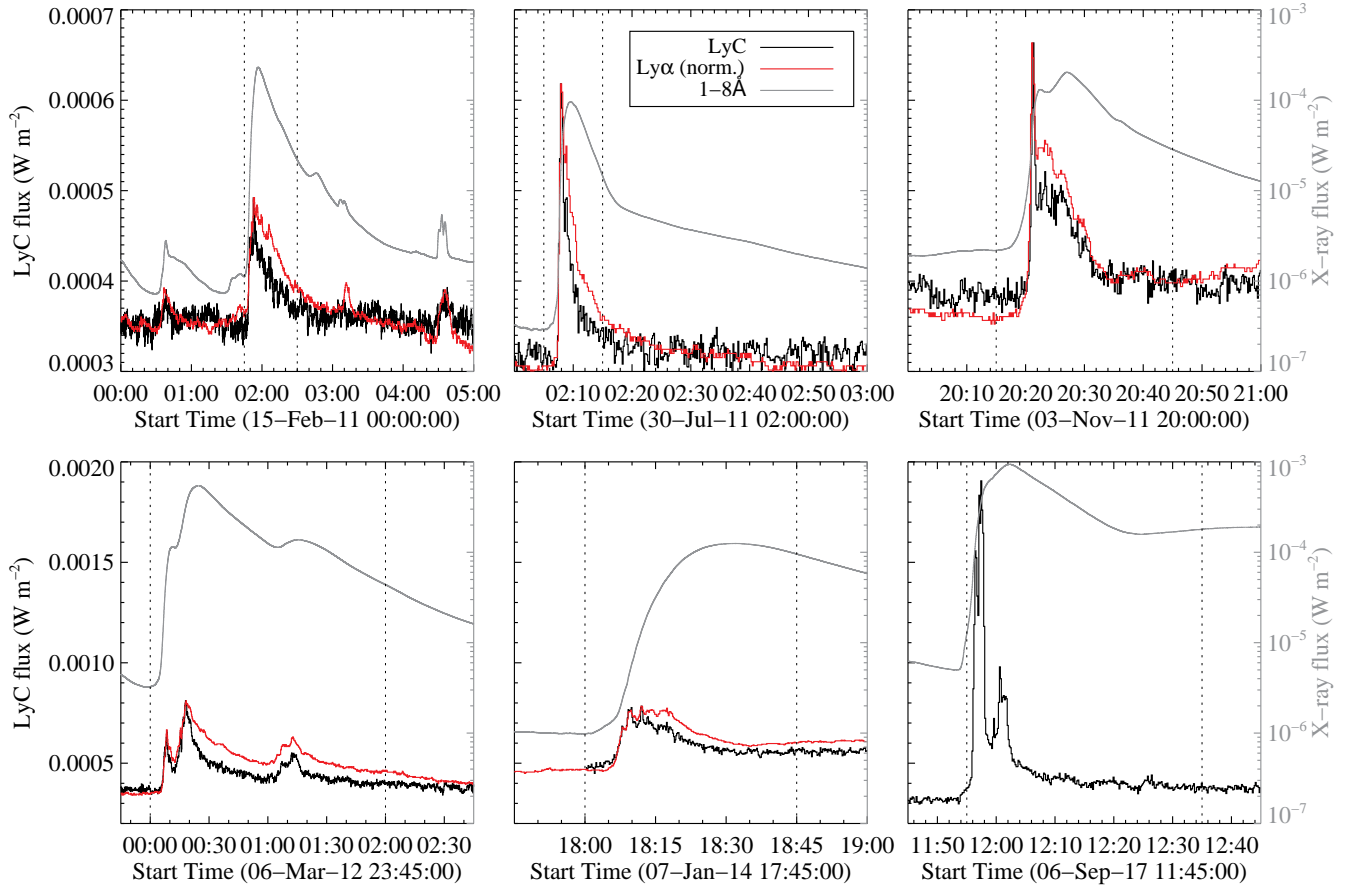}
\caption{LyC lightcurves (solid black curves) from SDO/EVE for each of the six events considered in this study. Also shown are the GOES/XRS (1--8~\AA; grey) and GOES/EUVS Ly$\alpha$ (1216~\AA) time profiles (red), which have been normalized to the LyC profiles. Note that the Ly$\alpha$ data for the 2017 September 6 flare had not been released at the time of writing. The vertical dotted lines in each panel denote the timerange over which the spectra were fitted to derive the $T_c$ and $b_1$ values.}
\label{fig:lya_lyc_sxr}
\end{center}
\end{figure*}

Figure~\ref{fig:lyc_head_ratio} shows two sample LyC spectra from SDO/EVE MEGS-B detected both before the 2017 September 6 X9.3 flare, and at the time of its peak impulsive phase emission. The spectra plotted in red and black in the bottom panel were taken at the times of the red and black vertical dashed lines in the top panel, respectively. Similarly, the blue and cyan lightcurves in the top panel are plotted for the wavelengths denoted by the vertical blue and cyan dashed lines in the bottom panel, respectively. The flux ratio at the head of the continuum (912~\AA) during the peak of the impulsive phase relative to its pre-flare value was measured to be 3.37, while for the flux at 800~\AA\ it was 4.78. This illustrates the flux increase and the change of the continuum slope as a function of $\lambda$ - a spectral hardening - indicative of an increase in color temperature, $T_c$. This event is the strongest and most complex among the six major flares that we considered for this study, and which are listed in Table~\ref{tab:flares}. In some cases the data reveals evidence for multiple episodes of energy release and subsequent heating, most notably those of 2012 March 6--7.

\begin{table}
\begin{center}
   \caption{Flares considered in this study.}
   \begin{tabular}{lcc}
   \hline
Event			&GOES peak time		&GOES class \\
   \hline
SOL2011-02-15	&01:56~UT	&X2.2	\\   
SOL2011-07-30 	&02:09~UT	&M9.3	\\
SOL2011-11-03 	&20:27~UT	&X1.9	\\
SOL2012-03-07 	&00:54~UT	&X5.4	\\
SOL2014-01-07 	&18:32~UT	&X1.2	\\
SOL2017-09-06 	&12:02~UT	&X9.3	\\
   \hline
   \end{tabular}
\label{tab:flares}
\end{center}
\end{table} 

The LyC lightcurves for each of the six events considered in this study are shown in black in Figure~\ref{fig:lya_lyc_sxr}, along with the corresponding GOES-15 soft X-ray (SXR; grey) and Ly$\alpha$ (normalized to the peak of LyC; red) profiles from the XRS \citep{hans96} and EUVS \citep{vier07} instruments, respectively. Ly$\alpha$ measurements from GOES were chosen over those from EVE due to issues with the Ly$\alpha$ measurements taken by the SDO/EVE MEGS-P Si diode detector (see \citealt{mill16}). Similar to MEGS-P, GOES/EUVS records the solar Ly$\alpha$ flux with a $\sim$10~s cadence in a broadband manner (i.e. no spectral information), but has much more extensive coverage and more accurate calibration. Version 4 of the GOES/EUVS data were used in this study while Version 6 of SDO/EVE MEGS-B data were used. The LyC lightcurves were generated using the technique of \cite{mill14}, who isolated the line emission from the underlying continuum using the RANdom SAmple Consensus (RANSAC) method of \cite{fisc81}. The continuum emission was fit with a simple power law for the purpose of tracking changes in intensity. However, in order to derive more physical parameters we fit the Lyman continuum data using Equation~\ref{eqn:planck} across three different wavelength ranges: 800--912~\AA, 870--912~\AA, and 800--870~\AA~ (ignoring major emission lines). We thus obtained the values for the color temperature ($T_c$) and departure coefficient ($b_1$) from each wavelength interval, although only the 800--912~\AA\ range was used to determine these parameters as a function of time throughout each event (see Section~\ref{sec:flare_results}).

As the EVE data are given in units of spectral irradiance or flux density ($F_\lambda$; W~m$^{-2}$~nm$^{-1}$) they first needed to be converted into units of specific intensity ($I_\lambda$; erg~cm$^{-2}$~s$^{-1}$~\AA$^{-1}$~sr$^{-1}$) in order to be fit with a blackbody function. Outside of flaring times EVE spectra are integrated over the entire solar disk, and so the corresponding solid angle, $\Omega_{QS}$, is simply:

\begin{equation}
\Omega_{\mathrm{QS}} = \frac{\pi R{_\odot}^2}{(1\mathrm{AU})^2}.
\label{eqn:rsun}
\end{equation}
\noindent
But during a solar flare the emission detected by EVE comes predominantly from a much more localized area. Based on previous studies of chromospheric flare ribbons \citep[e.g.][]{mill14} the flaring area, $A_\mathrm{flare}$, was taken to be 10$^{18}$~cm$^{2}$. However, estimates of flare ribbon areas for X-class flares have ranged from 10$^{16}$~cm$^{2}$ \citep{kruc11} to $>$10$^{19}$~cm$^{2}$ \citep{kaza17} before considering any possible filling factors. The resulting solid angle value is therefore:
\begin{equation}
\Omega_{\mathrm{flare}} = \frac{A_\mathrm{flare}}{(1\mathrm{AU})^2}.
\label{eqn:rflare}
\end{equation}
The conversion from flux density, $F_\lambda$, to specific intensity, $I_\lambda$, then becomes:

\begin{equation}
I_\lambda = \frac{100F_\lambda}{\Omega}.
\label{eqn:f2i}
\end{equation}
This will also include a negligible contribution from the underlying quiet-Sun. The factor of 100 arises from the fact that 1~W~m$^{-2}$~nm$^{-1}$ = 100~erg~cm$^{-2}$~s$^{-1}$~\AA$^{-1}$.

\section{RESULTS}
\label{sec:results}

\begin{table*}
\begin{center}
\caption{Quiet-Sun color temperatures and departure coefficients from fitting three separate LyC wavelength ranges.}\label{tab:t_b1_qs}
   \begin{tabular}{lccccccc}
   \hline 
	 
    & & \multicolumn{2}{c}{800--912~\AA}
    & \multicolumn{2}{c}{870--912~\AA} 
    & \multicolumn{2}{c}{800--870~\AA} \\
    Event & Pre-flare time (UT)
    & $T_c$ & $b_1$ 
    & $T_c$ & $b_1$ 
    & $T_c$ & $b_1$ \\
        
    \hline
SOL2011-02-15 & 01:00:02+33min&9303$~\pm~$65  &  996$~\pm~$124 & 8743$~\pm~$190  &  331$~\pm~$133 & 8708$~\pm~$116  &  277$~\pm~$72 \\
SOL2011-07-30 & 02:00:14+6min&8760$~\pm~$80  &  354$~\pm~$60 & 8030$~\pm~$190  &  67$~\pm~$31 & 8603$~\pm~$161  &  244$~\pm~$90 \\
SOL2011-11-03 & 20:00:09+17min&9173$~\pm~$72  &  624$~\pm~$88 & 8269$~\pm~$174  &  91$~\pm~$37 & 9480$~\pm~$156  &  1133$~\pm~$337 \\
SOL2012-03-06 & 23:00:03+50min& 9212$~\pm~$88  &  775$~\pm~$133 & 8346$~\pm~$223  &  126$~\pm~$65 & 8950$~\pm~$170  &  443$~\pm~$160 \\
SOL2014-01-07 & 18:00:14+5min&9651$~\pm~$65  &  1393$~\pm~$161 & 9148$~\pm~$185  &  560$~\pm~$200 & 9130$~\pm~$119  &  498$~\pm~$122 \\
SOL2017-09-06 & 11:00:47+33min&10536$~\pm~$142  &  8858$~\pm~$1879 & 8933$~\pm~$336  &  569$~\pm~$387 & 10494$~\pm~$279  &  8175$~\pm~$3551 \\
\hline
   \end{tabular}
\end{center}
\end{table*} 

\begin{figure*}[!t]
\begin{center}
\includegraphics[width=\textwidth]{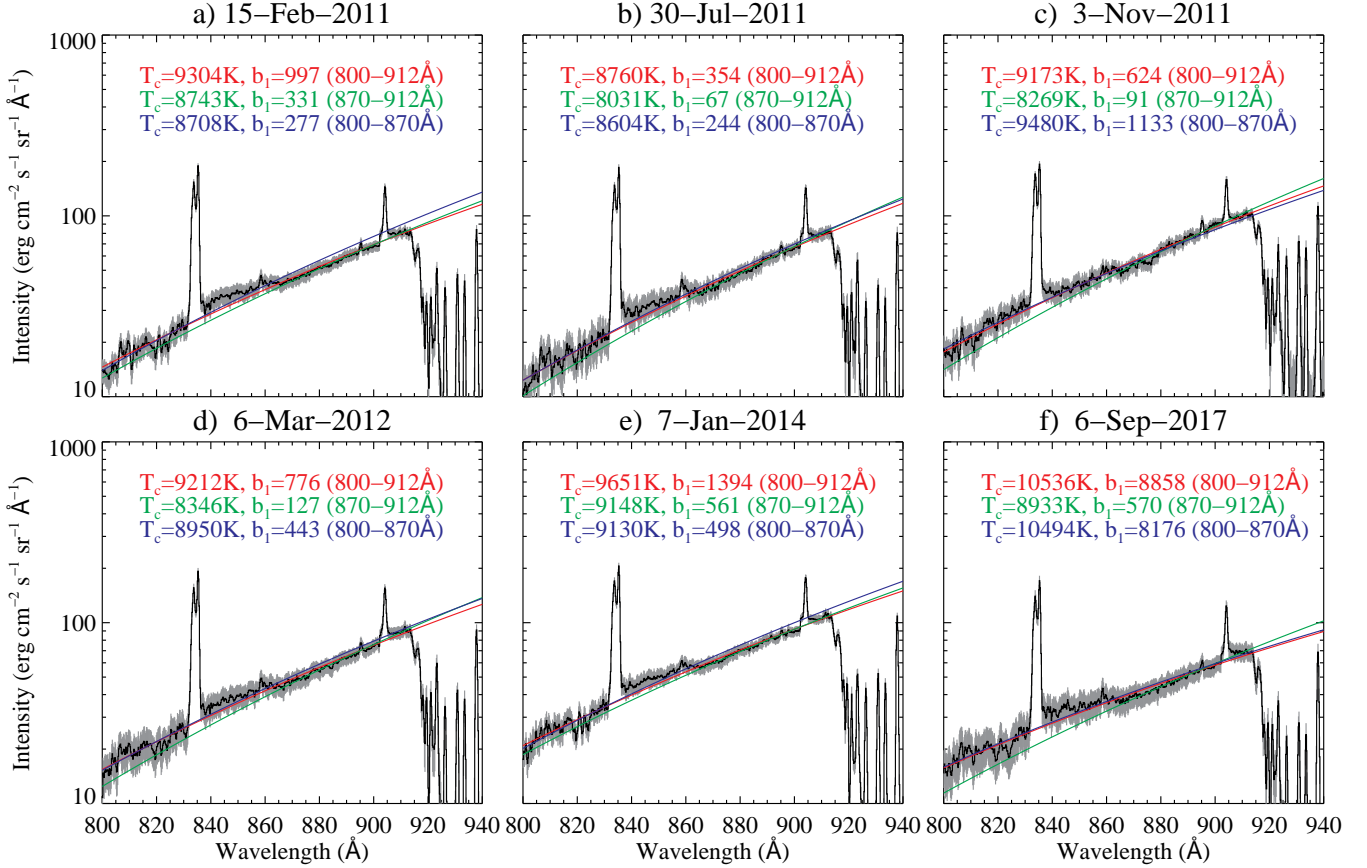}
\caption{EVE LyC spectra - in units of specific intensity - for quiescent periods prior to the onset of each flare considered in this study. The red, green, and blue curves are blackbody fits to the data (plus errors) over the 800--912~\AA, 870--912~\AA, and 800--870~\AA\ wavelength ranges, respectively. The derived $T_c$ and $b_1$ values for each fit interval are annotated on each panel.}
\label{fig:lyc_qs_fit}
\end{center}
\end{figure*}

\subsection{Quiet Sun}
\label{sec:qs_results}

Figure~\ref{fig:lyc_qs_fit} shows fits to LyC for quiescent periods prior to the onset of each of the six flares. After converting the EVE data into units of specific intensity, three different wavelength ranges were fit with a blackbody function: 800--912~\AA, 870--912~\AA, and 800--870~\AA. The values of $T_c$ and $b_1$ derived for each fit are annotated on each panel and are listed in Table \ref{tab:t_b1_qs} with associated uncertainties. Temperature values generally ranged from 8000--9500~K. These agree broadly with quiet-Sun values calculated by \cite{mach80} as well as with the values for the quiet-Sun and prominences reported by \cite{pare05}. It is likely that the temperature value for the 2017 September 6 flare (10500~K) may be higher than that of the others due to the presence of several large active regions on the visible solar disk at the time that contributed to the overall emission detected by EVE. Values for $b_1$ were typically on the order of 10$^2$--10$^3$ in line with those found for the quiet-Sun by \cite{mach80} and others.

\subsection{Flares}
\label{sec:flare_results}

\begin{table*}
\begin{center}
\caption{Flare impulsive phase color temperatures and departure coefficients from fitting three separate LyC wavelength ranges.}\label{tab:t_b1_flr}
   \begin{tabular}{lccccccc}
   \hline 
	 
    & & \multicolumn{2}{c}{800--912~\AA}
    & \multicolumn{2}{c}{870--912~\AA} 
    & \multicolumn{2}{c}{800--870~\AA} \\
    Event & LyC peak time (UT)
    & $T_c$ (K) & $b_1$ 
    & $T_c$ (K) & $b_1$ 
    & $T_c$ (K) & $b_1$ \\
        
    \hline

SOL2011-02-15 & 01:54:02~$\pm$25~s & 9880$~\pm~$278  &  0.7$~\pm~$0.3 & 9645$~\pm~$926  &  0.5$~\pm~$0.8 & 9305$~\pm~$493  &  0.2$~\pm~$0.2 \\
SOL2011-07-30 & 02:08:54~$\pm$25~s& 10692$~\pm~$276  &  1.8$~\pm~$0.7 & 9445$~\pm~$737  &  0.3$~\pm~$0.3 & 10750$~\pm~$554  &  2.0$~\pm~$1.6 \\
SOL2011-11-03 & 20:22:29~$\pm$25~s & 9425$~\pm~$289  &  0.3$~\pm~$0.1 & 9046$~\pm~$812  &  0.1$~\pm~$0.2 & 9744$~\pm~$630  &  0.5$~\pm~$0.5 \\
SOL2012-03-06 & 00:19:03~$\pm$25~s& 11316$~\pm~$138  &  1.7$~\pm~$0.3 & 10443$~\pm~$438  &  0.5$~\pm~$0.3 & 10793$~\pm~$244  &  0.8$~\pm~$0.3 \\
SOL2014-01-07 &18:13:04~$\pm$25~s& 11463$~\pm~$218  &  3.6$~\pm~$1.0 & 9500$~\pm~$553  &  0.2$~\pm~$0.2 & 11007$~\pm~$392  &  1.9$~\pm~$1.0 \\
SOL2017-09-06 &11:57:37~$\pm$25~s& 16607$~\pm~$110  &  68$~\pm~$5 & 12357$~\pm~$277  &  2.4$~\pm~$0.7 & 15923$~\pm~$188  &  43$~\pm~$6 \\

\hline
   \end{tabular}
\end{center}
\end{table*} 

\begin{figure*}[!t]
\begin{center}
\includegraphics[width=\textwidth]{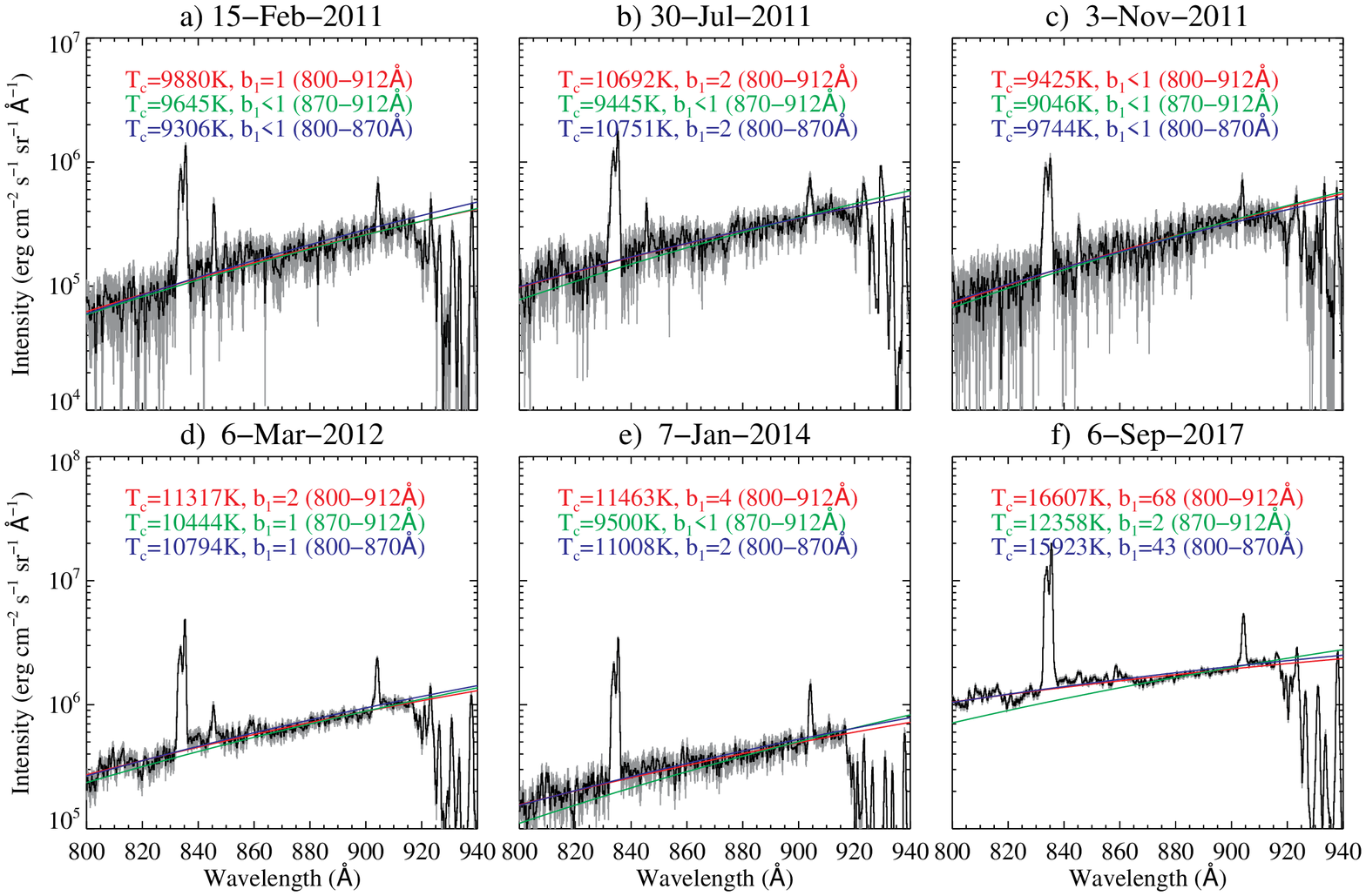}
\caption{Flare excess LyC spectra - in units of specific intensity - for the impulsive peaks of each flare considered in this study. The red, green, and blue curves are blackbody fits to the data (plus errors) over the 800--912~\AA, 870--912~\AA, and 800--870~\AA\ wavelength ranges, respectively. The derived $T_c$ and $b_1$ values for each fit interval are annotated on each panel.}
\label{fig:lyc_flr_fit}
\end{center}
\end{figure*}

The approach described in Section~\ref{sec:qs_results} was repeated for the EVE spectra around the time of the peak of the impulsive phase after subtracting out a pre-flare spectra (see Table~\ref{tab:t_b1_flr}). These data and their respective fits are shown in Figure~\ref{fig:lyc_flr_fit}. It can be seen that after subtracting a pre-flare spectrum and accounting for the change in emitting area, the LyC intensity has increased by factors of 10$^3$--10$^4$ relative to the pre-flare values (from around 10$^2$~erg~cm$^{-2}$~s$^{-1}$~sr$^{-1}$~\AA$^{-1}$ at the head of the continuum to 10$^5$--10$^6$~erg~cm$^{-2}$~s$^{-1}$~sr$^{-1}$~\AA$^{-1}$). This is in agreement with increases found by \cite{lema04}, as well as that predicted by \cite{ding97} for nonthermal electron beam fluxes of $\sim$10$^{12}$~\flux. The spectra are also notably harder than those of the quiet-Sun, thereby indicating a much higher color temperature. Values were typically in the range 10000--12500~K, with the 2017 September 6 flare showing values up to 17000~K. In all cases the value of $b_1$ dropped significantly relative to quiescent values, often approaching unity. In some cases $b_1<1$, which could, in principle, be due to effects other than thermal ionization, as shown in some numerical simulations and by \cite{ding97}. However, it is much more likely to be a consequence of the constant area that we arbitrarily assumed and we therefore cannot attribute any physical significance to this behavior.

\begin{figure*}[!t]
\centering
\includegraphics[width=0.92\textwidth]{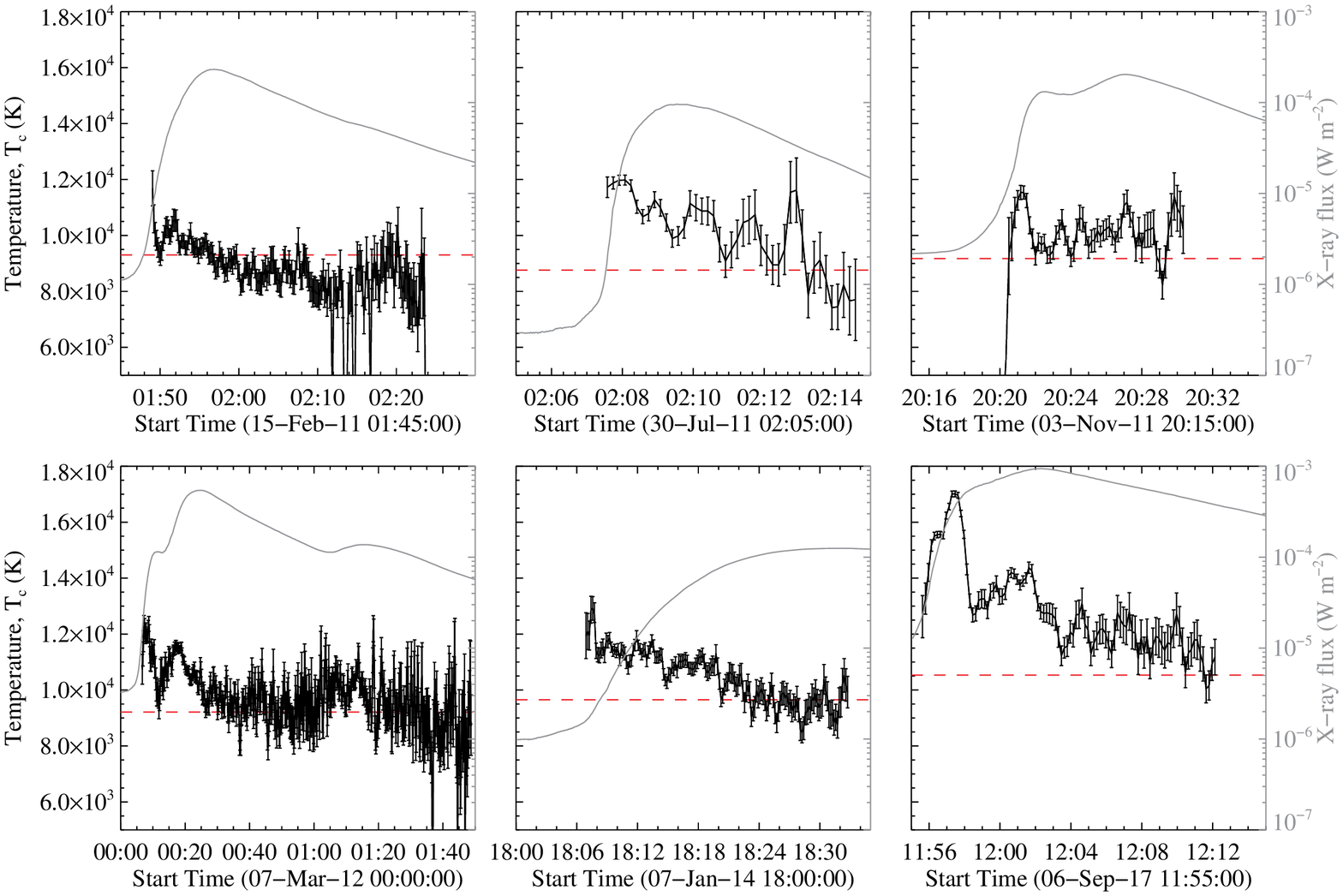}
\caption{Time profiles of $T_c$ for each of the six flares studied. The horizontal red dashed line in each panel denotes the quiet-Sun $T_c$ values presented in Figure~\ref{fig:lyc_qs_fit}.}\label{fig:coltemp}
\includegraphics[width=0.93\textwidth]{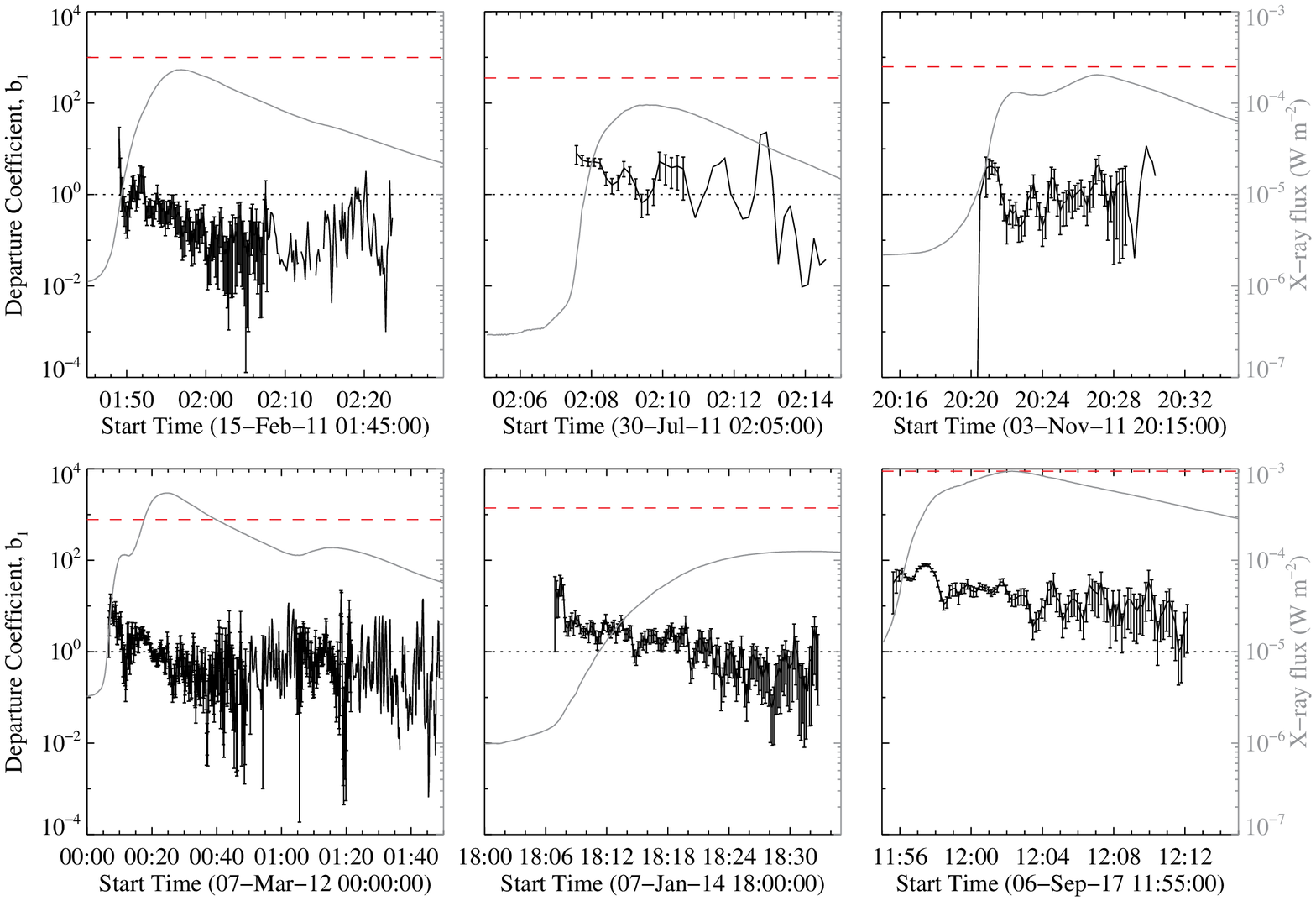}
\caption{Time profiles of $b_1$ for each of the six flares studied. The horizontal dotted line in each panel denotes the $b_1$=1 value, while the horizontal red dashed line denotes the quiet-Sun $b_1$ values presented in Figure~\ref{fig:lyc_qs_fit}.}\label{fig:b1}
\vspace{-0.48564pt}
\end{figure*}

The major advantage that EVE provides over previous LyC observations is that changes in $T_c$ and $b_1$ can now be tracked over time as a flare evolves thanks to its ability to provide spectra every 10~s. Figure~\ref{fig:coltemp} shows the changes in $T_c$ for each of the six events studied using only the fits to the 800--912~\AA\ range. In each case, around the flare onset when an excess spectrum becomes detectable, the color temperature is significantly (on the order of a few thousand Kelvin) above its quiet-Sun value (given by the red horizontal dashed lines). The flare then proceeds to cool beyond the impulsive phase once energy deposition has ceased.

In all cases our estimates show that the peak color temperature generally exceeds 10$^4$~K; significantly higher than the \cite{mach78} results, where it was determined as very similar to the average QS temperatures of $\sim$8500~K. The flare temperatures reported here agree more closely with those deduced from the lower wavelength end of the spectra (700--790\AA) that \cite{mach78} analyzed. As we can see in Figure~\ref{fig:coltemp} the $T_c$ estimate is well above QS values as soon as the early flare manifestations become evident. In many cases the temperature showed a significant drop before the rise phase of the flare had ended but it usually remained above 10$^4$~K. These higher temperatures may be due to the stronger flare classes of our  sample compared to that of \cite{mach78}.

Similarly, \cite{hein12} also found that they had to scale down the LyC intensity in the F2 model of \cite{mach80}, corresponding to a medium-to-strong flare, to match observations taken at the peak of the 2011 February 15 event (the first flare in our list in Table~\ref{tab:flares}). Since they did not calculate any $T_c$ value, although the LyC spectrum slope quite reasonably matched the F2 calculation, they concluded that the intensity mismatch was due to the fact that the flare area was rather small ($\sim$1.5$\times$10$^{18}$~cm$^2$) compared to typical X-class flares, but consistent with that of chromospheric kernels and white light flare (WLF) patches where the most energetic particles fluxes are presumed to precipitate.

Since EVE's Sun-as-a-star spectra does not give an indication of the area of the flaring regions we have ``normalized'' our analysis with the arbitrary assumption that the flare emission originated from a source with a canonical area of 10$^{18}$~cm$^2$; about half the size of a subflare in the H$\alpha$ class convention. Looking at the $b_1$ estimates in Figure~\ref{fig:b1}, we note that in many cases, particularly as time progresses, these turn out to be less than unity. This is an artifact due to the fixed area that we have assumed when converting from flux density to specific intensity, which implies that the observed flare excess spectrum originated in an area of 10$^{18}$~cm$^{2}$. In our methodology, $b_1$ scales linearly with the flare area. Therefore, if the actual flare areas were larger, particularly at the flare peak, the $b_1$ values would have to be larger for the fitted $B_\lambda(T)$ spectrum. Nevertheless, the departure coefficient would still be $b_1 \lesssim 10$ if we assumed a more typical flare ribbon area of $10^{19}$~cm$^2$ \citep{kaza17}.

On the contrary, it is also possible that if, during the impulsive phase, most of the emission were to originate in a small particle precipitation areas $<$10$^{18}$~cm$^2$, the estimated $b_1$ might result in somewhat smaller values to maintain the observed flux. These limitations indicate that the NLTE coefficients obtained from EVE data, or by the same token with any full Sun instrument, only give an order of magnitude estimate to what the actual value may have been. Although UV flare images, e.g. from SDO's Atmospheric Imaging Assembly (AIA; \citealt{leme12}), could be used to infer the flaring area and its evolution in time, we note that such images are heavily saturated for most of the duration of the events studied here. In spite of this drawback, compared with data obtained with e.g. the HCO-Skylab ATM instrument that had a 5$''$ spatial resolution, the fact that the $b_1$ estimate drops from values of the order of 10$^3$ in the QS records to close to unity in the observed flares, confirms the \cite{mach78} results showing that the emitted radiation originates in much higher density and presumably deeper atmospheric regions.

To present a more complete characterization of the LyC observational data, we decided to include Ly$\alpha$ line observations, noting that simultaneous observations of Ly$\alpha$ and LyC in flares are rarely reported in the literature. Ly$\alpha$ is a well observed feature in these events but is also a major single radiative energy loss spectral feature that also spans a wide range of formation altitudes from the base of the TR to the temperature minimum region (see \citealt{mach80,avre86,mill14,brow18}). As shown in Figure~\ref{fig:lya_lyc_sxr} both LyC and Ly$\alpha$ show a definite impulsive behavior, thus indicating that this phase is likely associated with the chromospheric heating due to accelerated particles in a thick-target fashion. In the LyC analysis this phase also shows a simultaneous increase in $T_c$ to $>$10$^4$~K, indicative of the presence of ambient electrons heated to such temperatures, and a low value of $b_1$, which even though it cannot be accurately estimated it becomes much lower than its pre-burst values and close to unity. These low $b_1$ values at the estimated temperatures are only compatible with the presence of a high-density medium where high collision frequencies induce a strong coupling of the continuum source function, $S_\lambda$, with the local Planck function, $B_\lambda(T)$, as shown in Equation~\ref{eqn:source}. In their model calculations, \cite{mach78}, \cite{mach80}, and \cite{avre86} have shown this to occur at electron densities $n_e \gtrsim 10^{13}$~cm$^{-3}$ - typical values obtained from chromospheric flare spectra and  at flaring chromospheric heights in radiative-hydrodynamic simulations \citep[e.g.][]{sves72,kerr16,simo17}.

\section{DISCUSSION}
\label{sec:disc}
Before going into the interpretation of these results, we highlight two main aspects seen in 
the $T_c$ and $b_1$ determinations. Firstly, in Figures~\ref{fig:lyc_qs_fit} and \ref{fig:lyc_flr_fit} we explicitly show the analysis of the pre-flare and impulsive phase LyC spectra of all flares, respectively, by fitting across three different wavelength intervals. These show significant increases in $T_c$ in all cases relative to the QS, and the corresponding decrease in $b_1$ obtained for a flare area of 10$^{18}$~cm$^2$. The QS values are in broad agreement with those of \cite{mach78}. However, the flare spectra consistently show values in excess of 10$^4$~K; larger than those reported by \cite{mach78} either because of the brighter flare classes (note that here also the largest $T_c$ corresponds to the strong X9.3 event on 2017 September 6) or the use of a wider wavelength interval in the determination.

Secondly, the increase of $T_c$ with decreasing wavelength reported by \cite{mach78} and asserted by VAL (see \citealt{avre08} for details), while not so apparent, and in some cases nonexistent in our brighter flare sample, may still be significant given the uncertainties (cf. Table~\ref{tab:t_b1_flr} and Figure~\ref{fig:lyc_flr_fit}). While this may be evidence for an optically thin layer formed at higher altitudes (see discussion below), we did not fit our spectra below 800\AA\ as \cite{mach78} did. This region of the spectrum becomes increasing populated by emission lines, which may lead to the formation of a pseudo-continuum, but may also have been recorded at a different time during a flare than at the head of the continuum due to the spectral scanning nature of the HCO instrument.

Having now established these common properties in the observed sample of events, it is appropriate to discuss their implications. The increase of $T_c$ and decrease of $b_1$ during flares are not surprising, considering that the flare-heated base of the transition region would have much larger electron densities due to the increased ionization fraction of hydrogen. From this, two effects follow: the increase of the collisional rate causing the decrease in the NLTE departure coefficient, $b_1$, and the significant increase in the LyC opacity, subsequently formed in a deeper region with a higher temperature.

Assuming characteristic conditions of the flaring atmosphere, we can estimate the thickness $\Delta z$ of the $\tau_{LyC}$ = 1 layer using the optical depth of LyC, given by
\begin{equation}
\tau_{LyC} = \sigma_1 n_1 \Delta z.
\end{equation}
Here we assume a photo-ionization cross-section $\sigma_1 = 1.05 \times 10^{-26} \lambda^3$ cm$^{-2}$, with large hydrogen ($n_H$) and electron densities ($n_e$), both of order 10$^{12-13}$~cm$^{-3}$ in the temperature regime around 10$^4$~K (see e.g \citealt{mach80}'s Table 4 and \citealt{avre86}). Considering that the population of the first hydrogen level, $n_1$ is in the 10$^{11-12}$~cm$^{-3}$ range \citep[e.g.][]{mach80}, the thickness of the formation layer of the LyC, $\Delta z$, will be at most a few hundred kilometers, if not smaller, with a temperature range of $\lesssim$10$^{4}$~K to 2$\times$10$^{4}$~K; exactly the range of values found in our flare sample. The dependence of the photo-ionization cross-section on the wavelength ($\sigma_1 \propto \lambda^3$) would only give a difference of $\approx 2$ between 700~\AA\ to 912~\AA. This simple calculation indicates that under flaring conditions the formation layer of the LyC will easily be optically thick, independent of wavelength. Moreover, all our estimates for $b_1$ during flares indicate that the plasma approaches LTE conditions, so that $S_\lambda \simeq B_\lambda(T)$, and therefore $T_c \simeq T_e$. We thus conclude that contrary to the QS situation,  where the LyC formation depth spans a somewhat extended region \citep{avre08}, in our flare sample the LyC is formed in a rather narrow shell with characteristic temperatures around $10^4$~K, close to LTE conditions, $b_1 \approx 1$. 

In an electron beam heating scenario, these depths are such that only electrons with energies $\gtrsim$100~keV can directly reach the LyC forming layer, and by the same token, that of the Lyman lines. It follows that the lower energy electrons that carry the bulk of the beam energy, deposit it at higher altitudes and lower mass column densities that span the range between 5$\times$10$^{-6}$~g~cm$^{-2}$, at the top of the quiescent pre-flare chromosphere, and 10$^{-3}$~g~cm$^{-2}$, where material previously at chromospheric temperatures is impulsively heated to $T>10^6$~K. This X-ray emitting plasma is thus transparent to the hydrogen spectrum. 

Only one of the flares in our sample (2017 September 6) exhibits a higher $T_c$ value measured at shorter wavelengths than at the continuum head, similar to \cite{mach78}. This could indicate a contribution from an optically thin layer with a higher temperature to the formation of the LyC. Given that the corresponding $b_1$ value is significantly greater than unity also supports this possibility (see Table~\ref{tab:t_b1_flr}). 

While the hydrogen ionization fraction becomes large above 20,000~K, which leads to a steep decrease in the overall hydrogen opacity, the column depth of neutral hydrogen, $n_1 \Delta z$, could still be sufficiently large to enhance the LyC optical depth, $\tau_{LyC}$. Since the increase of $B_\lambda(T)$ between temperatures of the order of 8500~K and a few times 10$^4$~K is larger at 700~\AA\ than at 912~\AA, any significant contribution from an optically thin layer would reinforce the continuum more strongly at the shorter wavelengths. Evidently, this optically thin layer cannot extend into the corona, where the high temperatures and low densities prevent the recombination of protons and electrons into neutral hydrogen, effectively making $\tau_{LyC}$ negligibly small. That imposes a limit on both the geometrical thickness, $\Delta z$, of this hypothetical optically thin layer and to its temperature, $T$. This contribution from a hotter layer to the color temperature at shorter wavelengths is a possibility that needs to be investigated with detailed radiative hydrodynamic simulations.

As noted by VAL and by \cite{avre08}, the observed QS brightness temperature of LyC increases with decreasing wavelength, not only in the HCO-Skylab spectrometer observations but also hinted in more recent SUMER observations as well \citep{curd01}. This result is counterintuitive since the LyC opacity decreases with decreasing wavelength ($\sigma_1 \propto \lambda^3$) and therefore unit optical depth at, say, 700~\AA\ {\it should} occur deeper in the atmosphere than the continuum head at 912~\AA. Since in QS, as well as in flare models, the temperature and presumably the continuum source function decrease with depth, one should intuitively observe a decrease of the color temperature at shorter wavelengths, contrary to what is observed. \cite{avre08} show, through the analysis of the height dependence of the contribution functions at various wavelengths, that for the shorter wavelengths, there are not only contributions from the chromospheric layers but also from the high-temperature layers of the transition region, which have less influence on the intensity for the wavelengths at the head of the Lyman continuum. This contribution from the hotter layers must originate from an optically thin region or else the color temperature would simply reflect the higher temperature instead. As shown in Equation~\ref{eqn:source}, the continuum source function, $S_\lambda$, is approximately the Planck function divided by the departure coefficient, $b_1$, which measures the hydrogen level-1 departures from LTE. As mentioned above, the increase of $B_\lambda(T)$, and hence $S_{\lambda}$, between temperatures of the order of 8500~K and a few times 10$^4$~K is larger at 700~\AA\ than at 912~\AA. Therefore, in order to match the observations the temperature must increase abruptly to high values just above the region where the head of the Lyman continuum is formed. \cite{avre08} show that this is the situation in the QS, and give a detailed analysis of the reasons for the abrupt temperature increase in the models that we recommend to read. The contribution from an optically thin layer would suggest that {\em for shorter wavelengths} the condition to estimate the electron temperature $T_e$ directly from the color temperature $T_c$, i.e. $T_c \simeq T_e$, is not met.

\section{CONCLUSIONS}
\label{sec:conc}

In this study we have performed a quantitative analysis of the observed response of Lyman continuum emission during six major solar flares. The findings presented show a common pattern in its behavior that, independent of any flare modeling assumption, indicate how the lower solar atmosphere responds to impulsive energy release. Our analysis shows that the impulsive phase is characterized by a strong and rapid increase in LyC intensity, which is associated with a spectrum with a higher color temperature ($T_c$ $\gtrsim$ 10$^4$~K) and a strong coupling of the continuum source function with $B_\lambda(T)$. This indicates that LyC emission, when coming from a stratified atmosphere, originates in a relatively thin ($\Delta z \lesssim$ 100~km), deep, dense, layer ($m> 10^{-3}$~g~cm$^{-2}$ or equivalently $N>10^{21}$~cm$^{-2}$), with electron densities in excess of 10$^{13}$~cm$^{-3}$. 

A dynamic response analysis of this phase is beyond the scope of our analysis, but we can speculate that part of this material is compressed and pushed downwards, creating the observed high temperature flare ribbons, while the hottest component explosively evaporates into the coronal soft X-ray loops (e.g., \citealt{mill09}). What is clearly important is the evidence that the energy balance in the whole flaring atmosphere may not be a straightforward cause and effect relationship, such as the direct particle heating of the overall LyC and Lyman lines in the flaring chromosphere. Moreover, we note that although direct electron beam heating seems to strongly affect the ionization of \ion{He}{1} to \ion{He}{2} \citep{simo16}, this process regulates the local plasma temperature, which in turn may affect the overall conditions for thermal conduction. 

This work emphasizes the value of studying continuum emission processes in the flaring chromosphere. \cite{ding97} claim that comparisons with model predictions of LyC emission can be used to infer the properties of nonthermal electrons (if present) in the absence of HXR or microwave observations. Tentative evidence for enhancements of Balmer and Paschen continua during flares have also recently been presented by \cite{hein14} and \cite{kerr14}, respectively. Joint observations of each of these hydrogen continua would be a powerful tool in furthering our understanding of the issue of energy transport during solar flares.

\acknowledgments
The authors would like to thank the anonymous referee for their constructive feedback that greatly improved the quality of this paper. We would also like to thank Prof. Lyndsay Fletcher and Dr. Graham Kerr for many insightful discussions. ROM would like to acknowledge support from NASA LWS/SDO Data Analysis grant NNX14AE07G, and the Science and Technologies Facilities Council for the award of an Ernest Rutherford Fellowship (ST/N004981/1). P.J.A.S. acknowledges support from the University of Glasgow's Lord Kelvin Adam Smith Leadership Fellowship. The research leading to these results has received funding from the European Community's Seventh Framework Programme (FP7/2007--2013) under grant agreement No. 606862 (F-CHROMA).

Sadly, Marcos Machado passed away just as we were preparing to submit the final draft of this manuscript. Marcos was a pioneer of solar flare physics at the dawn of space-based solar physics research, particularly with regard to the flaring solar chromosphere, for which he generated state-of-the-art semi-empirical models in the 1970's and 80's. This paper marked a return to one that he published in 1978 on Lyman continuum emission during solar flares, and one which he was excited to see updated with more advanced data from SDO. It was an honor and a privilege for the both of us to work with Marcos on a topic so dear to him, and for him to impart his wisdom and experience upon us with kindness, humor, and a generosity of spirit.

\end{document}